\documentclass[a4paper,conference]{IEEEtran}
\usepackage{cite}
\usepackage{amsmath,amssymb,amsfonts,amsthm}
\usepackage{algorithm}
\usepackage{algorithmic}
\usepackage{graphicx}
\usepackage{textcomp}
\usepackage{xcolor}
\usepackage{bbm}
\usepackage{nccmath}
\usepackage{booktabs}
\usepackage{mathtools}
\usepackage{subfig}

\def\BibTeX{{\rm B\kern-.05em{\sc i\kern-.025em b}\kern-.08em
    T\kern-.1667em\lower.7ex\hbox{E}\kern-.125emX}}

\usepackage{tikz}
\usetikzlibrary{shapes.arrows}
\usetikzlibrary{plotmarks}
\usetikzlibrary{patterns}
\usetikzlibrary{arrows.meta}
\usetikzlibrary{positioning}

\usepackage{pgfplots}
\usepgfplotslibrary{colorbrewer}
\usepackage{adjustbox}

\usepackage{gincltex}

\pgfplotsset{compat=1.15}

\usepackage[acronym]{glossaries}

\newacronym{aoi}{AoI}{Age of Information}
\newacronym{paoi}{PAoI}{Peak AoI}
\newacronym{sic}{SIC}{successive interference cancellation}
\newacronym{bs}{BS}{base station}
\newacronym{pmf}{pmf}{probability mass function}
\newacronym{cdf}{CDF}{Cumulative Distribution Function}
\newacronym{irsa}{IRSA}{irregular repetition slotted ALOHA}
\newacronym{sinr}{SINR}{signal-to-interference-plus-noise ratio}
\newacronym{snr}{SNR}{signal-to-noise ratio}
\newacronym{kpi}{KPI}{Key Performance Indicator}
\newacronym{plr}{PLR}{packet loss rate}
\newacronym{csi}{CSI}{channel state information}
\newacronym{ofdma}{OFDMA}{Orthogonal Frequency-Division Multiple Access}
\newacronym{oma}{OMA}{Orthogonal Multiple Access}
\newacronym{noma}{NOMA}{Non-Orthogonal Multiple Access}
\newacronym{pnoma}{PNOMA}{Partial Non-Orthogonal Multiple Access}
\newacronym{rlnc}{RLNC}{Random Linear Network Coding}
\newacronym{mds}{MDS}{Maximum Distance Separable}
\newacronym{urllc}{URLLC}{ultra-reliable low-latency communications}
\newacronym{embb}{eMBB}{enhanced mobile broadband}
\newacronym{mmtc}{mMTC}{massive machine-type communications}
\newacronym{bec}{BEC}{binary erasure channel}
\newacronym{ran}{RAN}{radio access network}
\newacronym{lr}{LR}{latency-reliability}

\theoremstyle{definition}
\newtheorem{definition}{Definition}

\title{\vspace{0.05cm}Slicing a single wireless collision channel among throughput- and timeliness-sensitive services}

\author{\IEEEauthorblockN{Israel Leyva-Mayorga, Federico Chiariotti, \v Cedomir Stefanovi\' c, Anders E. Kal{\o}r,  and Petar Popovski}
\IEEEauthorblockA{Department of Electronic Systems, Aalborg University\\
Fredrik Bajers Vej 7C, 9100 Aalborg, Denmark, email: \{ilm,fchi,cs,aek,petarp\}@es.aau.dk\vspace{-0.4cm}}
}

\begin{document}

\maketitle
\begin{abstract}
The fifth generation (5G) wireless system has a platform-driven approach, aiming to support heterogeneous connections with very diverse requirements. The shared wireless resources should be sliced in a way that each user perceives that its requirement has been met. Heterogeneity challenges the traditional notion of resource efficiency, as the resource usage has cater for, e.g. rate maximization for one user and \emph{timeliness} requirement for another user. This paper treats a model for \gls{ran} uplink, where a throughput-demanding broadband user  shares wireless resources with an intermittently active user that wants to optimize the timeliness, expressed in terms of latency-reliability or \gls{aoi}. We evaluate the trade-offs between throughput and timeliness for \gls{oma} as well as \gls{noma} with \gls{sic}.
We observe that \gls{noma} with \gls{sic}, in a conservative scenario with destructive collisions, is just slightly inferior to that of \gls{oma}, which indicates that it may offer significant benefits in practical deployments where the capture effect is frequently encountered.
On the other hand, finding the optimal configuration of \gls{noma} with \gls{sic} depends on the activity pattern of the intermittent user, to which \gls{oma} is insensitive. 

\end{abstract}

\glsresetall
\section{Introduction}

Perhaps the main innovation in 5G is its platform approach to connectivity, aiming to support any type of connection through a suitable combination of three generic connection types with vastly different requirements: \gls{embb}, \gls{urllc}, and \gls{mmtc}~\cite{3GPPTR38913}. 
\gls{embb} aims at delivering enhanced data-rates in comparison to 4G. \gls{urllc} usually involve exchange of small data amounts at a very low latency (say, $1$\,ms) and very high reliability (e.g. packet loss rate $10^{-5}$). Finally, \gls{mmtc} aims at supporting sporadic transmission of small data chunks, but from a vast number (e.g. thousands) of devices in a single cell. 

\emph{Network slicing} refers to the allocation 
of the network resources among the active services, such that the network is able to provide performance guarantees and meet their widely different requirements~\cite{Popovski2018}. 
When related to the slicing of the shared wireless resource in a \gls{ran} context, we term it wireless slicing. It has been studied in the form of diverse \gls{oma} and \gls{noma} techniques in the presence of multiple users with the same service type~\cite{Maatouk2019, Dai2015, Wu2018} and 
the trade-offs in achievable data rates for \gls{embb} services are clearly characterized~\cite{Dai2015, Wu2018}.
However, further research is needed on novel slicing mechanisms for heterogeneous services to support the widely diverse traffic patterns and requirements of \gls{urllc}, \gls{mmtc}, and \gls{embb}~\cite{Richart2016}.
In this respect, a common uplink scenario involves two categories of users: broadband and intermittent users.
Broadband users transmit data continuously, aiming to maximize the throughput.
In contrast, intermittent users transmit short packets sporadically, being mainly interested in the \emph{timeliness} of their data. 
\gls{urllc} and \gls{mmtc} users fall in the latter category, having stringent and relaxed timeliness requirements, respectively.

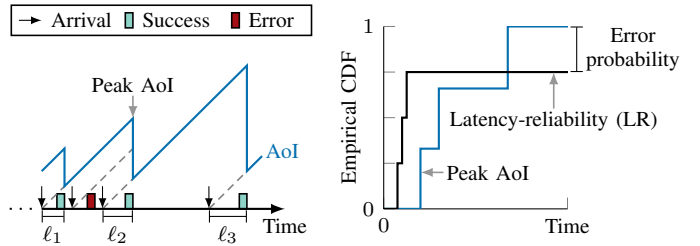
\begin{figure}[t]
\captionsetup{belowskip=-0.5cm}
\centering
\vspace{0.1cm}
\begin{tikzpicture}[>=latex, scale=1,font=\footnotesize]

\pgfmathsetmacro{\pone}{0.3}
\pgfmathsetmacro{\ptwo}{0.7}
\pgfmathsetmacro{\pthree}{1.2}
\pgfmathsetmacro{\pfour}{2.7}
\pgfmathsetmacro{\pw}{0.1}
\pgfmathsetmacro{\ph}{0.2}
\pgfmathsetmacro{\paone}{0.3}
\pgfmathsetmacro{\patwo}{0.3}
\pgfmathsetmacro{\pathree}{0.4}
\pgfmathsetmacro{\pafour}{0.5}

\filldraw[semithick,fill=Set3-A] (\pone-\pw,0) rectangle (\pone,\ph);
\filldraw[semithick,fill=Reds-K] (\ptwo-\pw,0) rectangle (\ptwo,\ph);
\filldraw[semithick,fill=Set3-A] (\pthree-\pw,0) rectangle (\pthree,\ph);
\filldraw[semithick,fill=Set3-A] (\pfour-\pw,0) rectangle (\pfour,\ph);

\draw[->](\pone-\paone,0.3)--++(0,-0.3);
\draw[->](\ptwo-\patwo,0.3)--++(0,-0.3);
\draw[->](\pthree-\pathree,0.3)--++(0,-0.3);
\draw[->](\pfour-\pafour,0.3)--++(0,-0.3);

\draw[densely dashed, semithick, gray](\pone-\paone,0)--++(\paone,\paone);
\draw[densely dashed, semithick, gray](\ptwo-\patwo,0)--++(\pthree-\ptwo+\patwo,\pthree+\patwo-\ptwo);
\draw[densely dashed, semithick, gray](\pthree-\pathree,0)--++(\pathree,\pathree);
\draw[densely dashed, semithick, gray](\pfour-\pafour,0)--++(\pafour,\pafour);

\draw[thick,->] (0,0)node[anchor=east, inner sep=1pt]{$\dotsc$}--(\pfour+0.5,0)node[below,font=\footnotesize]{Time}; 

\draw[thick, PuBu-J] (0,0.5)--(\pone,\pone+0.5)--(\pone,\paone)--++(\pthree-\pone,\pthree-\pone)--(\pthree,\pathree)--++(\pfour-\pthree,\pfour-\pthree)--(\pfour,\pafour)--++(0.2,0.2)node[right, inner sep=1pt, yshift=2pt,font=\footnotesize]{AoI};

\draw[|-|] (\pone-\paone,-0.1)--++(\paone,0)node[below, midway, font=\footnotesize]{$\ell_1$};
\draw[|-|] (\pthree-\pathree,-0.1)--++(\pathree,0)node[below, midway, font=\footnotesize]{$\ell_2$};
\draw[|-|] (\pfour-\pafour,-0.1)--++(\pafour,0)node[below, midway, font=\footnotesize]{$\ell_3$};

\draw[Greys-G,<-, semithick] (\pthree,\paone+\pthree-\pone)--++(0,0.3)node[above, inner sep=1pt, black,font=\footnotesize]{Peak AoI};

\pgfmathsetmacro{\ly}{2.5}
\pgfmathsetmacro{\ldx}{1.4}
\pgfmathsetmacro{\xd}{-0.3}

\draw[->](\xd,\ly)--++(0.3,0)node[anchor=west,font=\footnotesize]{Arrival};
\filldraw[semithick,fill=Set3-A] (\ldx+\xd,\ly-0.5*\ph) rectangle (\ldx+\pw+\xd,\ly+0.5*\ph);
\node[anchor=west, align=center,font=\footnotesize] at (\ldx+\pw+\xd,\ly){Success};
\filldraw[semithick,fill=Reds-K] (2*\ldx+\xd,\ly-0.5*\ph) rectangle (2*\ldx+\pw+\xd,\ly+0.5*\ph);
\node[anchor=west, align=center,font=\footnotesize] at (2*\ldx+\pw+\xd,\ly){Error};

\draw[semithick] (\xd-0.1,\ly-0.2) rectangle (\pfour+0.85,\ly+0.2);

\begin{scope}[xshift=4.5cm]
\pgfmathsetmacro{\xmx}{4}
\begin{axis}[
axis lines*=left,
height=4cm,
width=4cm,
ymin=0,
ymax=1,
ytick={0,0.25,0.5,0.75,1},
yticklabels={$0$,,,,$1$},
xmin=0,
xmax=\xmx,
xtick={0,\xmx},
xticklabels={$0$,Time},
ylabel={Empirical CDF},
ylabel shift = -6 pt,
clip mode=individual
]
\addplot[PuBu-J, const plot, thick] coordinates{(-1,0)(\pone+0.5,0.33)(\pthree,0.66)(\pfour,1)(\xmx+1,1)};
\addplot[black, const plot, thick] coordinates{(-1,0)(\paone,0.25)(\pathree,0.5)(\pafour,0.75)(\xmx+1,0.75)};
\draw[Greys-G,<-, semithick] (axis cs: \pone+0.5,0.2)--(axis cs: \pone+1,0.2)node[right, inner sep=1pt, black,font=\footnotesize]{Peak AoI}; 
\draw[Greys-G,<-, semithick] (axis cs: 3.7,0.75)--(axis cs:3.7,0.55)node[below, inner sep=1pt, black,font=\footnotesize, align=center]{Latency-reliability (LR)};
\draw[|-|](\xmx+0.2,0.75)--(\xmx+0.2,1)node[midway,right, align=center]{Error\\probability};
\end{axis}
\end{scope}
\end{tikzpicture}
\vspace{-0.3cm}
\caption{Exemplary diagram of the \gls{aoi} and \gls{lr} \glspl{kpi} in a period with four packet transmissions. The latency for packets transmitted with errors is set to $\infty$.}
\label{fig:aoi_vs_lr_diag}
\end{figure}

For the intermittent users with timeliness requirements, there are two different \glspl{kpi} that may be relevant.
The first is the \gls{aoi}, which represents the time elapsed since the generation of the last received packet~\cite{Kaul2012}, and is useful when the intermittent user sends updates of an ongoing process.
In this case, the information freshness is more important than any single update and, thus, packet loss is tolerable, while the update generation pattern is a key factor to ensure a low age~\cite{Kosta2020}.
In applications requiring high reliability, the high percentiles of the \gls{paoi}, defined as the maximum \gls{aoi} achieved immediately before the reception of an update~\cite{Costa2014}, are the most relevant metric.
The other \gls{kpi} is a combination of latency and reliability, useful for \gls{urllc}-like applications that send critical data.
We exploit a \gls{lr} metric measuring the time between the generation and successful reception of each packet in the form of a \gls{cdf}~\cite{Nielsen2018}, assuming that lost packets have infinite latency, and using high percentiles of the \gls{cdf} as a \gls{kpi}. 
Fig.~\ref{fig:aoi_vs_lr_diag} illustrates the difference between \gls{paoi} and LR during an observation period with four packet transmissions. 

Non-orthogonal network slicing has been investigated in combination with \gls{sic} in uplink scenarios with heterogeneous services. 
For instance, \cite{Popovski2018} investigated the benefits of \gls{noma} and \gls{oma} assuming that (i) \gls{embb} users are allocated orthogonal resources among them, (ii) there is a single \gls{urllc} user, and (iii) \gls{mmtc} traffic is Poisson distributed.
It was observed that \gls{noma} may offer benefits with respect to (w.r.t.) \gls{oma} depending on the rate of the \gls{embb} users and whether the coexisting traffic is \gls{urllc} or \gls{mmtc}.
The work was later extended to a multi-cell scenario with strict latency guarantees for \gls{urllc} traffic~\cite{Kassab2019}: a single \gls{urllc} user per cell was assumed and it was observed that \gls{noma} leads to a greater spectral efficiency w.r.t. \gls{oma}.
A similar conclusion was drawn by Maatouk \emph{et al.}~\cite{Maatouk2019} in an uplink scenario where two intermittent users aim to minimize the \gls{aoi}. 
Their results show that a greater spectral efficiency does not directly translate into a lower average \gls{aoi}. 

In this paper, we investigate the trade-offs and performance bounds with orthogonal and non-orthogonal slicing with \gls{sic} for the case where a single frequency band is sliced to accommodate one broadband and one intermittent user. To the best of our knowledge, this is the first paper to explore the inherent differences between slicing the radio access for \gls{lr}- and for \gls{aoi}-oriented services with additional broadband traffic.
In particular, we provide closed-form expressions for the achievable throughput (given in packets per slot) for the broadband user and the timeliness for the intermittent user.
Our analysis in this baseline scenario \emph{(i)} illustrates the fundamental performance trade-offs between access methods; \emph{(ii)} can be directly extended to multiple frequency bands and a greater number of users; and \emph{(iii)} provides the basis to explore complex slicing approaches.
The scenario assumed in the paper is a conservative one, based on a simple model in which collision are destructive -- no involved packet can be directly decoded (e.g., through capture effect).
Consequently, the results presented in the paper may be assumed to correspond to lower bounds on the performance of \gls{noma}.
We have observed that \gls{noma} greatly outperforms \gls{oma} when the intermittent user requires \gls{lr}, but is not able to maintain a message (i.e., update) queue. 
Additionally, \gls{noma} can outperform \gls{oma} with queuing if the target is extremely low latency or extremely high throughput.
On the other hand, the achievable \gls{paoi} with \gls{oma} is considerably lower than that with \gls{noma}, as \gls{oma} is able to tolerate higher arrival rates with a higher reliability.
These results highlight the need to adapt the system to the particular timeliness requirement posed by the intermittent user. 

In the rest of the paper, Section~\ref{sec:system_model} presents the general system model, while we derive the distribution of the \glspl{kpi} for \gls{oma} and \gls{noma} in Section~\ref{sec:analyses}.
Section~\ref{sec:results} discusses simulation results, while Section~\ref{sec:concl} concludes the paper and lists possible avenues for future work.

\section{System model}
\label{sec:system_model}

We consider a general uplink scenario where multiple users transmit data to a \gls{bs}.
A time-slotted and frequency-division multiple access channel is considered.
Our focus is on \emph{one} of the many possible frequency bands (subcarriers) where a broadband and an intermittent user are allocated.
The broadband user falls under the full-buffer traffic model, where messages of length of $K$ packets are encoded into $N$ linearly independent coded packets using a packet-level coding scheme. This can be achieved with, e.g., \gls{mds} codes and infinite queue of the broadband user.
Changing the coding rate $K/N$ can trade off rate for robustness against packet erasures, caused by noise errors, but also collisions with the packets from an intermittent user.
The intermittent user generates new messages at each time slot with probability $\alpha$.
Each message fits into a single packet, which the intermittent users transmits just once at the next available slot (without application of a packet-level code).
The intermittent user can maintain a queue of up to $Q$ of the generated packets. If a new packet arrives when the queue is full, the oldest packet is discarded and the new packet is kept. 

\begin{figure}[t]
\captionsetup{belowskip=-0.5cm}
	\centering
        \includegraphics{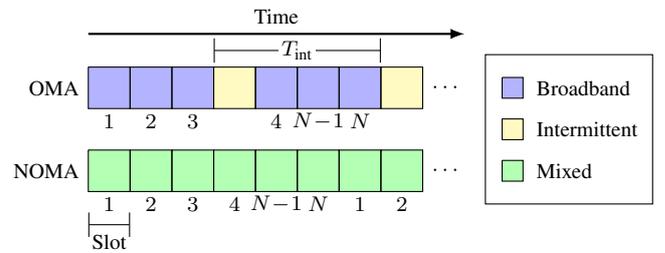}
 \caption{Frame structure for \gls{oma} and \gls{noma}; $K=4$, $N=6$.}
 \label{fig_system}
\end{figure}

The \gls{bs} allocates resources to the users. We define $\mathcal{U}=\{1,2\}$, where user $1$ is the broadband user and user $2$ is the intermittent user, and assume three different allocations of a particular time slot: 1) Broadband, allocated to user $1$; 2) Intermittent, allocated to user $2$; and 3) Mixed, allocated to both users. We define the following two access methods. \emph{(1) \gls{oma}} Only broadband and intermittent resources are allocated. $T_\text{int}$ is the period between intermittent resources. The broadband user is not affected by the activity of the intermittent user. \emph{(2) \gls{noma}}: Only mixed resources are allocated. Hence, the frame structure is exclusively determined by $K$ and $N$. Here the rate and reliability of the broadband user are dependent on the level of activity of the intermittent user.
The frame structures for these methods are illustrated in Fig.~\ref{fig_system}.

Finally, we consider a \gls{bec} with collisions (in \gls{noma} case) and constant transmission power at each user.
The erasure probabilities for the broadband and intermittent user are denoted as $\epsilon_1$ and $\epsilon_2$, respectively; these depend on the received \gls{snr} and the receiver sensitivity.
 The considered collision model is simple, where the \gls{sinr} is below the receiver sensitivity every time that users $1$ and $2$ transmit at the same resource.
 That is, a packet may only be immediately decoded at a particular slot if only one user transmits.
 The \gls{bs} stores collisions in a frame and may able to decode transmissions from the intermittent user \emph{after} decoding the block of $K$ packets from the broadband user.
 For this, implements \gls{sic} to remove the interference from the broadband user.
 Then, the packet from the intermittent user is decoded (at the expense of increased latency) if the \gls{snr} is sufficient. 



\section{Performance analysis}
\label{sec:analyses}

In this section, we derive the \glspl{kpi} for the \gls{oma} and \gls{noma} systems, for a \gls{lr}- or \gls{paoi}-oriented intermittent user with activation probability $\alpha$.
In the following, we define the multinomial function $\text{Mult}(\mathbf{K};N,\mathbf{p})$ as
\begin{IEEEeqnarray}{c}
    \text{Mult}(\mathbf{K};N,\mathbf{p})=\frac{N!\prod_{i=1}^{|\mathbf{p}|}p_i^{K_i}(1-\sum_{i=1}^{|\mathbf{p}|}p_i)^{N-\sum_{i=1}^{|\mathbf{p}|}K_i} }{(N-\sum_{i=1}^{|\mathbf{p}|}K_i)!\prod_{i=1}^{|\mathbf{p}|} K_i!},\IEEEeqnarraynumspace
\end{IEEEeqnarray}
where $|\mathbf{p}|$ is the length of vectors $\mathbf{p}=\left[p_1,p_2,\dotsc,p_{|\mathbf{p}|}\right]$ and $\mathbf{K}=\left[K_1,K_2,\dotsc,K_{|\mathbf{p}|}\right]$. The binomial function $\text{Bin}(K;N,p)$ is the special case with \mbox{$|\mathbf{p}|=1$.}

\subsection{LR-oriented OMA}
In \gls{oma} the intermittent user has a reserved slot every $T_{\text{int}}$.
Denoting the decoding success probability of the broadband user as $p_{s,1}$, the expected throughput $S_1$ is
\begin{align}
  S_1&=p_{s,1}\frac{(T_{\text{int}}-1)K}{T_{\text{int}}N}\\
  p_{s,1}&=\sum_{r=K}^N \text{Bin}(r:N,1-\epsilon_1).
\end{align}
As the broadband user can only use $T_{\text{int}}-1$ slots for each $T_{\text{int}}$, setting up more frequent transmission opportunities for the intermittent user will reduce the throughput. 

To derive the latency \gls{pmf} for the intermittent user, we must compute the state-transition probabilities $\mathbf{P}^{(0)}=\left[P^{(0)}_{ij}\right]$ of its queue, given as
\begin{equation}
  P^{(0)}_{ij}=\begin{cases}
           \text{Bin}(q_j-q_i+1;T_{\text{int}},\alpha)&\text{if } q_i\leq q_j+1<Q; \\
           \sum_{m=Q-q_i}^{T_{\text{int}}} \text{Bin}(m;T_{\text{int}},\alpha)&\text{if } q_j=Q-1,
         \end{cases}\label{eq:trans_queue}
\end{equation}
where \mbox{$q_i,q_j\in\{0,1,\dotsc,Q\}$} are the queue states before and after a transmission, respectively.
The steady-state distribution of the queue immediately after a transmission \mbox{$\boldsymbol{\pi}^{(0)}=\left[\pi^{(0)}_0,\pi^{(0)}_1,\dotsc,\pi^{(0)}_Q\right]$} is the left-eigenvector of $P^{(q)}$ with eigenvalue $1$, normalized to sum to $1$ to be a valid probability measure. That is,
\begin{equation}
 \boldsymbol{\pi}^{(0)}(\mathbf{I}-\mathbf{P}^{(0)})=0;\qquad\text{s.t.} \sum_{q=0}^Q\pi^{(0)}_q=1.
 \label{eq:steady_mat}
\end{equation}
We calculate the steady-state distribution of the queue occupation $n$ slots after an intermittent user transmission as
\begin{equation}
  \pi_q^{(n)}=\begin{cases}
              \sum_{s=0}^q \pi^{(0)}_s \text{Bin}(q-s;n\alpha)&\text{if } q<Q;\\
              \sum_{s=0}^Q\sum_{m=Q-s}^n \pi^{(0)}_s \text{Bin}(m;n,\alpha)&\text{if } q=Q.
            \end{cases}\label{eq:steady_slot_oma}
\end{equation}
If a packet is queued behind $q$ others, it will be transmitted at the $q+1$-th opportunity, unless new arrivals make the system drop some of the packets ahead of it in the queue -- if the queue is full, the oldest packet (i.e., the first in the queue) is dropped. Let $\mathcal{G}_{\ell}^{(n)}=\{0,\ldots,T_{\text{int}}-n\}\times\{0,\ldots,T_{\text{int}}\}^{\ell-1}$ be the set containing the possible numbers of new packets generated by the intermittent user in the next $\ell$ transmission windows of length $T_\text{int}$ after the considered packet is generated in the $n$ slots after the last intermittent slot.
The probability of each element $\mathbf{g}\in\mathcal{G}_{\ell}^{(n)}$ is simply:
\begin{equation}
  p_{\mathcal{G}_{\ell}^{(n)}}(\mathbf{g})=\text{Bin}(g_1;T_{\text{int}}-n,\alpha)\prod_{i=1}^\ell \text{Bin}(g_i;T_{\text{int}},\alpha).
\end{equation}
At each transmission opportunity, one packet is transmitted, and other packets are dropped if the number of generated packets exceeds the number of remaining places in the queue. For a given generation vector $\mathbf{g}\in\mathcal{G}_{\ell}^{(n)}$, the considered packet is transmitted at the $\ell$-th transmission opportunity, where $\ell$ is the first index that satisfies 
\begin{equation}
  \psi_k^{(\mathbf{g},q)}=\delta\left(\sum_{i=1}^k\left[q+1-Q+\sum_{j=1}^i g_j\right]^+\!+k-(q+1)\right),
\end{equation}
where $\delta(x)$ is the delta function, equal to $1$ if $x=0$ and $0$ otherwise, and $[x]^+=\max(x,0)$.
We now define the set
\begin{equation}
  \mathcal{S}_{\ell}^{(n,q)}=\Big\{\mathbf{g}\in\mathcal{G}_{\ell}^{(n)}: \psi_\ell^{(\mathbf{g},q)}-\sum_{k=1}^{\ell-1}\psi_k^{(\mathbf{g},q)}=1\Big\},
\end{equation}
with the elements $\mathbf{g}\in\mathcal{G}_{\ell}^{(n)}$ for which the considered packet is transmitted at the $\ell$-th opportunity.
Since the packet is either transmitted within $q+1$ transmission attempts or discarded, the success probability for the intermittent user is given by
\begin{equation}
  p_{s,2}=\sum_{n=1}^{T_{\text{int}}}\sum_{q=0}^{Q}\sum_{\ell=1}^{q+1}\pi^{(n-1)}_q\sum_{\mathclap{\mathbf{g}\in\mathcal{S}_{\ell}^{(n,q)}}}p_{\mathcal{G}_{\ell}^{(n)}}(\mathbf{g})\frac{(1-\epsilon_2)}{T_{\text{int}}}.
\end{equation}
Knowing that packet generation probability is the same for every slot, we now use \eqref{eq:steady_slot_oma} to uncondition the latency \gls{pmf}
\begin{equation}
 p_T(\ell T_{\text{int}}-n)=\sum_{n=1}^{T_{\text{int}}}\sum_{q=0}^{Q}\frac{\pi^{(n-1)}_q \sum_{\mathbf{g}\in\mathcal{S}_{\ell}^{(n,\min(q,Q-1))}}p_{\mathcal{G}_{\ell}^{(n)}}(\mathbf{g})}{T_{\text{int}} p_{s,2}(n,\min(q,Q-1))}.
\end{equation}

\subsection{PAoI-oriented OMA}

Preemption is the optimal strategy to minimize \gls{paoi}.
Therefore, we set $Q=1$ for the \gls{paoi}-oriented \gls{oma} and packets are always sent at the first available transmission opportunity. 
Hence, the \gls{pmf} of the delay of a successful transmission is
\begin{equation}
  p_{T}(t)=\frac{\alpha(1-\alpha)^t}{1-(1-\alpha)^{T_{\text{int}}}}.
\end{equation}
We now compute the \gls{paoi} for the \gls{oma} scheme, which is given by the sum of the transmission delay and the inter-transmission time, denoted by $Z$. Since exactly one slot every $T_{\text{int}}$ is reserved for user 2, $Z$ is equal to  $T_{\text{int}}$ times the number of reserved slots between consecutive transmissions. The latter is a geometric random variable, whose parameter \mbox{$\xi=(1-(1-\alpha)^{T_{\text{int}}})(1-\epsilon_2)$} is the probability of a successful transmission. Hence, we obtain the \gls{pmf} of the \gls{paoi} $\Delta$ as 
\begin{equation}
 p_{\Delta}(T_{\text{int}}z+t)=(1-\xi)^{z-1}\xi p_{T}(t).
\end{equation}

\subsection{LR-oriented NOMA}
In the \gls{noma} case all slots are mixed and the queue size is irrelevant; any intermittent packet is immediately transmitted and queued at the receiver until \gls{sic} is performed. Hence, we set $Q=1$ and define $p_1=1-\epsilon_1$ and $p_2=\alpha(1-\epsilon_2)$; the latter is the probability that, at each slot after \gls{sic}, an intermittent packet is transmitted and received correctly. 

As a starting point, we derive the success probability and throughput of user 1, respectively, as
\begin{align}
    p_{s,1}&=\sum_{r_2=0}^{N-K}\text{Bin}(r_2;N,\alpha)\sum_{r_1=K}^{N-r_2}\text{Bin}(r_1;N-r_2,p_1),\\
S_1&=\frac{K}{N}p_{s,1}.
\end{align}
As intermittent packets are transmitted immediately, we simply need to compute the delay $T$ between the slot in which the packet is transmitted and when it is decoded. First, we define $F$ to be the random variable of the slot within the frame where the \emph{first decoding event} (for the intermittent user) occurs. As a result of the simple collision model, the first decoding event can only occur after the broadband user is decoded. Therefore, the \gls{pmf} of $F$, whose support is $\{K+1,\ldots,N\}$, is given as
\begin{equation}
\label{eq:F}
    p_F(f)=\sum_{r_2=1}^{f-K}\sum_{r_1=K}^{f-r_2}\text{Mult}((r_1,r_2);f,(p_1,p_2))-\sum_{i=1}^{f-1}p_F(i).
\end{equation}
 It is easy to compute the \gls{cdf} of $F$, denoted $P_F(f)$. The probability of having at least one successful decoding in a frame is then $P_F(N)$, so we get $p_{s,2}=P_F(N)$.
The probability of receiving $R$ more packets from user $2$ in a frame, conditioned on the first decoding event $F$, is 
\begin{equation}
 p_R(r;f)=\frac{\text{Bin}(k;N,p_2)}{P_F(N)}.
\end{equation}
We  compute the probability of having a decoding event $D$ in a given slot, conditioned on $F$, as
\begin{equation}
\label{eq:D}
 p_D(d;f)=\begin{cases}
               \sum_{r=0}^{N-d}p_R(r;d) &\text{if } f=d;\\
               \sum_{r=0}^{N-f}\frac{p_R(r;f)r}{(N-f)} &\text{if }f>d.
              \end{cases}
\end{equation}
Combining \eqref{eq:F} and \eqref{eq:D} yields
\begin{equation}
 p_D(d)=\sum_{f=K+1}^d p_F(f)p_D(d;f).\label{eq:delivered}
\end{equation}
In order to compute the \gls{pmf} of the delay $T$, we consider three cases. First, the delay is always 0 if the decoding event is not the first in the slot, as \gls{sic} can be performed immediately. Second, the first decoding event has delay 0 if the broadband user frame was already decoded before slot $d$, but no packets from user 2 were successfully received before $d$. Hence,
\begin{equation}
    p_{T|F}(0;d)=\sum_{\mathclap{r_1=K}}^{d-1}\frac{\text{Mult}((r_1,0);d-1,((1-\alpha)p_1,p_2))p_2}{p_F(d)}.
\end{equation}
Third, the delay for the first decoding event is larger than 0 if there were intermittent packets before slot $d$, but the broadband user frame was not decoded yet. In this latter case, we also need to consider that multiple packets might be decoded in the first decoding event. Therefore, we compute the conditional \gls{pmf} of the delay of the last packet $\ell$ given that $c$ packets are decoded in the first decoding event as
\begin{equation}
    p_{T_\ell|F,C}(t;d,c)=\frac{c (d-t)! (d-c)!}{ (d-1)! (d-t-c)!}.\label{eq:lastlatency}
\end{equation}
Let $T_p$ be the the delay of the previous packets, whose \gls{pmf} is
\begin{equation}
    p_{T_p|F,C,T_\ell}(t;d,c,t_\ell)=\frac{c-1}{d-(t_\ell+1)},\,t\in\{t_\ell+1,\ldots,d\}.
\end{equation}
The overall delay \gls{pmf} for a given number of simultaneous decoded packets $c$ is
\begin{equation}
\label{eq:T}
p_{T|F,C}(t;d,c)=p_{T_\ell|F,C}(t;d,c)+\sum_{t_\ell=0}^{t-1}p_{T_p|F,C,T_\ell}(t;d,c,t_\ell).
\end{equation}
Further, we calculate the probability of decoding $C$ packets simultaneously if the first decoding event is in slot $d$ as
\begin{equation}
    p_C(c;d)=\frac{\text{Mult}((K-1,c);d-1,(p_1,p_2))p_1}{p_F(d)}.
\end{equation}
The latter allows us to uncondition \eqref{eq:T} on $C$ and compute the overall delay for the first decoding event in the reliable transmission scenario as 
\begin{equation}
    p_{T|F}(t;d)=\sum_{c=1}^{d-K}p(T|F,C)(t;d,c)p_C(c;d).
\end{equation}
Finally, the probability that a received packet is part of the first decoding event is given by
\begin{equation}
    p_{F|D}(d)=\sum_{c=1}^{d-K}\sum_{r=0}^{N-d}\frac{p_F(d)p_C(c;d)p_R(r;d)c}{(c+r)p_D(d)P_F(N)},
\end{equation}
and the overall delay \gls{pmf} is
\begin{equation}
    p_T(t)=\!\begin{cases}
    \ \ \ \sum\limits_{\mathclap{d=K+1}}^N p_D(d)p_{F|D}(d)p_{T|F}(t;d)&t>0;\\
    \ \ \ \sum\limits_{\mathclap{d=K+1}}^N p_D(d)(1-p_{F|D}(d)(1-p_{T|F}(0;d)))\!&t=0.
    \end{cases}\label{eq:delay_dist}
\end{equation}

\subsection{PAoI-oriented NOMA}

Most of the analysis from the \gls{lr}-oriented case is valid for the \gls{paoi}-oriented case, the only difference being that the only meaningful packet in a decoding event is the most recent one:
\begin{equation}
    p_{T|F,C}(t;d,c)=p_{T_\ell|F,C}(t;d,c),
\end{equation}
where $p_{T_\ell|F,C}(t;d,c)$ is the same as in~\eqref{eq:lastlatency}. This affects the probability of a packet being in the first decoding event in the frame, as computation no longer considers previous packets:
\begin{equation}
    p_{F|D}(d)=\sum_{r=0}^{N-d}\frac{p_F(d)p_R(r;d)}{(r+1)p_D(d)P_F(N)}.
\end{equation}
We compute the conditioned delay \gls{pmf} $p_T(t;d)$ by using the values in~\eqref{eq:delay_dist}. First, we compute the probability that a decoding event is the last in the frame as
\begin{equation}
 p_L(d)\!=\!\sum_{\mathclap{f=K+1}}^d p_F(f)\sum_{r=1}^{\mathclap{d-f}} \frac{p_D(d;f)(d\!-\!f\!-\!1)!(N\!-\!f\!-\!r\!+\!1)!}{(d\!-\!f\!-\!r\!+\!1)!(N\!-\!f\!-\!1)! p_D(d)},
\end{equation}
Next, let $\bar{L}$ indicate that the decoding event is not the last in the frame, i.e., that there are more decoding events in the frame after slot $d$.
Building on this, the \gls{pmf} of $Z$ when the decoding event is not the last in the frame
\begin{equation}
 p_{Z}(z;d,\bar{L})=\frac{(1-\alpha(1-\epsilon_2)^{z-1}\alpha(1-\epsilon_2)}{1-(1-\alpha(1-\epsilon_2)^{N-d}}
 \label{eq:z_sameframe}
\end{equation}
If the decoding event is the last in the frame, the inter-transmission time $Z$ is formed by three components: the $N-d$ slots until the end of the current frame, $e$ full frames, and the time $F$ until the first decoding event from the beginning of the frame.
Hence, we derive the \gls{pmf} of $Z$ in this case as
\begin{equation}
    p_{Z}((e-1)N+d+f;d,L)=(1-P_N(F))^{e}p_F(f).\label{eq:z_nextframe}
\end{equation}
Using~\eqref{eq:z_sameframe} and~\eqref{eq:z_nextframe} to uncondition on $L$, we get
\begin{equation}
 p_{Z}(z;d)=\begin{cases}
                                     p_{Z}(z;d,\bar{L})(1-p_L(d)) &z<N-d;\\
                                     p_{Z}(z;d,L)p_L(d) &z\geq N-d.
                                    \end{cases}\label{eq:z_uncond}
\end{equation}
With the \glspl{pmf} of $T$ and $Z$ from \eqref{eq:delay_dist} and~\eqref{eq:z_uncond}, along with $p_D(d)$ from \eqref{eq:delivered}, we can finally compute the \gls{pmf} of the \gls{paoi} by convolving $T$ and $Z$ and unconditioning over $D$. That is,
\begin{equation}
    p_{\Delta}(\tau)=\sum_{d=K+1}^N p_D(d)\sum_{t=0}^{\min(d,\tau)} p_{T}(t;d)p_{Z}(\tau-t;d).
\end{equation}


\section{Evaluation}\label{sec:results}

In this section, we perform the evaluation of the analyzed schemes in terms of the following \glspl{kpi}.
The throughput is the main \gls{kpi} for the broadband user, assuming that $K>1$.
The \glspl{kpi} of the intermittent user are: 1) the
$90$th percentile of the \gls{paoi} $\Delta_{90}=\max_{n}\{n : \Pr\left[\Delta<n\right]<0.9\}$,
and 2) the $90$th percentile of \gls{lr} $L_{90}=\max_{n}\{n : \Pr\left[L<n\right]<0.9\}$. 

As these \glspl{kpi} are interlinked, we evaluate their trade-offs for a specific activation probability $\alpha$ and erasure probabilities $\epsilon_1$ and $\epsilon_2$, via the \emph{Pareto frontier} defined in the following. 
\begin{definition}
Let $f:\mathbb{Z}^2\rightarrow \mathbb{R}^2$ and $\mathcal{C}$ be the set of feasible combinations of the parameter settings. Next, let \begin{equation}
    Y=\{(S_1,\tau):(S_1,\tau)=f(c\in\mathcal{C})\},
\end{equation}
where $S_1$ is the throughput of the broadband user and $\tau$ is the timeliness of the intermittent user (i.e., $\Delta_{90}$ or $L_{90}$). The \emph{Pareto frontier} $\mathcal{P}$ for a combination of $\alpha$, $\epsilon_1$, and $\epsilon_2$ is the set 
\begin{equation}
\mathcal{P}(Y)=\{(S_1,\tau)\in Y:\nexists(S'_1,\tau')\in Y:S_1<S'_1, \tau>\tau'\}.
\end{equation}
\end{definition}

To evaluate the evolution of the performance of each scheme with $\alpha$, we set a minimum throughput for the broadband user $S_{1_\text{min}}$.
Then, for each considered value of $\alpha$, we obtain the configuration of both \gls{oma} and \gls{noma} that results in the minimum timeliness for the intermittent user while maintaining $S_1\geq S_{1_\text{min}}$ for a given combination of $\epsilon_1$ and $\epsilon_2$; we call this the optimal configuration.
The parameter settings considered for the evaluation are listed in Table~\ref{tab:params}. 
All presented results were verified with Monte Carlo simulations, not shown in the plots as the simulated curves mirror the analytical results.

\begin{table}[t]
    \centering
    \captionsetup{belowskip=0.16cm}
    \caption{Parameter settings}
    \begin{tabular}{@{}lll@{}}
    \toprule 
    Parameter & Symbol & Setting\\\midrule
    User 1 source block size & $K$ & $\{2,\dotsc,64\}$\\
       User 1 coded block size  &  $N$ & $\geq K$\\
       User 1/User 2 erasure prob.& $\epsilon_1/ \epsilon_2$ & $0.1/0.05$\\
       User 2 activation probability & $\alpha$ & $\left[10^{-4},10^{-1}\right]$\\
      Intermittent slot period & $T_\text{int}$ & $\{1,\dotsc,64\}$\\
       Max. queue length (\gls{oma})&$Q$ & $\{1,4\}$\\
         \bottomrule
    \end{tabular}
    \label{tab:params}
    \vspace{-0.16cm}
\end{table}


Our first observation is that, for \gls{oma}, the values of $K$ and $N$ can be chosen independently of $\alpha$ and $T_\text{int}$ with the single objective of maximizing the throughput of the broadband user $S_1$. Afterwards, by carefully choosing $T_\text{int}$, the trade-off can be adjusted to ensure $S_1\geq S_{1_\text{min}}$ is achieved with the minimal \gls{paoi} or \gls{lr}.
This is due to the orthogonality between the two users, where the settings for user 1 and user 2 do not affect each other.
Hereafter, we denote the optimal values for $K$ and $N$ as $K^*$ and $N^*$. By fixing $K\leq64$, we got $K^*=64$, and $N^*=77$, which we use in the following.

\begin{figure}[t]
\captionsetup{belowskip=-0.5cm}
    \centering
    \includegraphics[width=0.97\columnwidth]{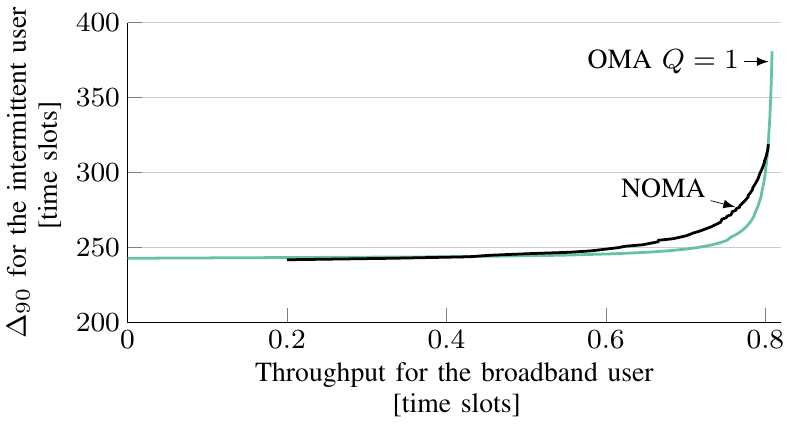}
    \caption{Pareto frontier for the throughput of user 1 $S_1$ vs. $90$th percentile of the \gls{paoi} of user 2 $\Delta_{90}$ for $\alpha=0.01$.}
    \label{fig:Pareto_front_oma_paoi}
\end{figure}

Next, we investigate the performance of heterogeneous access with a \gls{paoi}-oriented service.
Note that, in \gls{oma} case, queuing is not needed at the intermittent user (i.e., $Q=1$), as newly arrived packets supersede the older ones.
Fig.~\ref{fig:Pareto_front_oma_paoi} shows the Pareto frontier for $S_1$ vs. $\Delta_{90}$.
As it can be seen, the maximum $S_1$ is around $0.8$ with both \gls{oma} and \gls{noma}, with \gls{oma} achieving a slightly higher $S_1$ at the expense of a considerable increase in $\Delta_{90}$. Furthermore, $\Delta_{90}$ with \gls{oma} is relatively constant until $S_1\approx0.75$, after which it sharply increases. Conversely, the increase in $\Delta_{90}$ with \gls{noma} is clearly observed from $S_1\approx0.5$ and reaches a maximum of $\Delta_{90}\approx310$; this is lower than the maximum with \gls{oma}. In summary, \gls{oma} provides a better trade-off than \gls{noma} when $S_1$ is between $0.6$ and $0.8$. If the target is a greater or lower $S_1$, both access methods deliver comparable results. 

\begin{figure}[t]
\captionsetup{belowskip=-0.5cm}
    \centering
    \includegraphics[width=0.97\columnwidth]{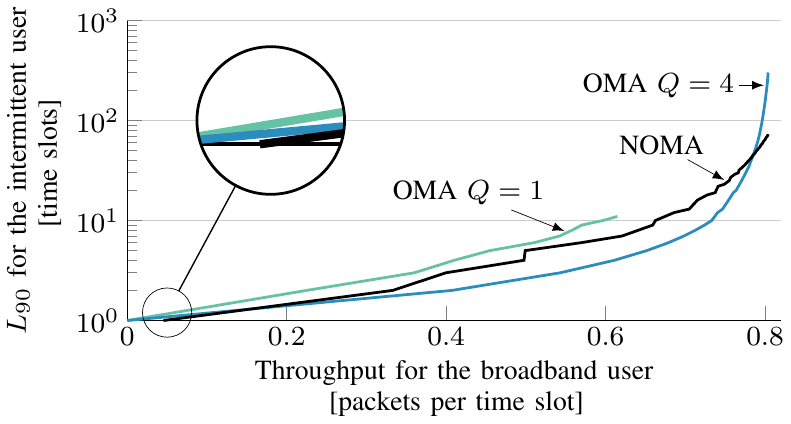}
    \caption{Pareto frontier for the throughput of user 1 $S_1$ vs. $90$th percentile of \gls{lr} for user 2 $L_{90}$ for $\alpha=0.01$.}
    \label{fig:Pareto_front_oma_l90_inv}
\end{figure}

The performance of heterogeneous access with a LR-oriented service is presented in Fig.~\ref{fig:Pareto_front_oma_l90_inv}, showing the Pareto frontier for $S_1$ vs. $L_{90}$.
In this case, queuing at the intermittent user is needed with \gls{oma} to achieve comparable results as with \gls{noma}.
This can clearly observed in Fig.~\ref{fig:Pareto_front_oma_l90_inv}: for \gls{oma} and $Q=1$, the maximum $S_1$ that can be achieved with $L_{90}<\infty$ is around $0.6$; and, for $Q=4$, the achievable $S_1$ grows to nearly $0.8$.
Further, \gls{oma} and \gls{noma} deliver similar values of $L_{90}$ for low values of $S_1$, and, while \gls{oma} results in a lower $L_{90}$ for $S_1\in\left[0.35,0.75\right]$, \gls{noma} delivers a lower $L_{90}$ for the maximum achievable $S_1$. 

\begin{figure}
\captionsetup{belowskip=-0.5cm}
    \centering
    \subfloat[]{\includegraphics{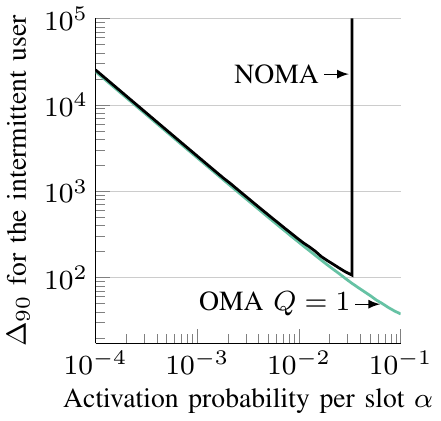}}
    \subfloat[]{\includegraphics{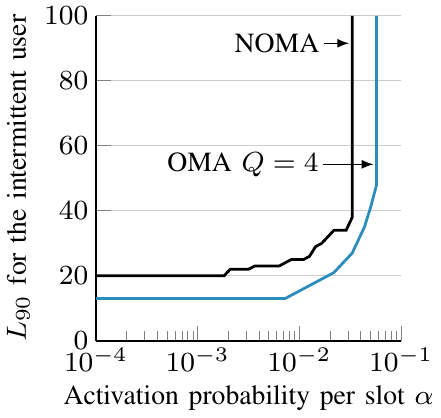}}\vspace{-0.2cm}\\
      \subfloat[]{\includegraphics{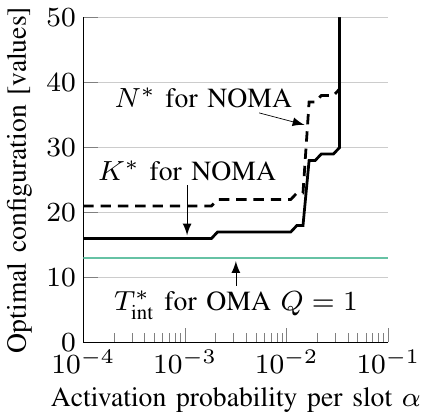}}
    \subfloat[]{\includegraphics{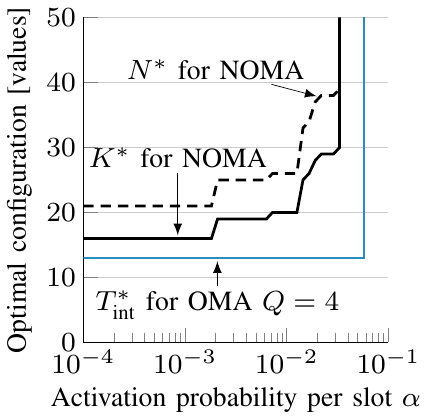}}
    \caption{Achievable timeliness for user 2 while ensuring $S_1\geq 0.75$: (a) $\Delta_{90}$ and (b) $L_{90}$, and optimal configuration of the access methods for (c) $\Delta_{90}$ and (d) $L_{90}$ as a function of $\alpha$.}
    \label{fig:stress}
\end{figure}

Finally, Fig.~\ref{fig:stress} shows the achievable timeliness for user 2 while ensuring $S_1\geq 0.75$ for a wide range of values of $\alpha$. There is an asymptote at the points where the requirements cannot be met. As it can be seen, \gls{oma} leads to a slightly lower $\Delta_{90}$ and $L_{90}$ than \gls{noma} for all values of $\alpha$. Besides, Fig.~\ref{fig:stress} exhibits that the configuration of \gls{oma} is trivial: select the optimal $K^*$ and $N^*$ for user 1 and, then, select the shortest possible $T_\text{int}$ to ensure $S_1\geq 0.75$ and the probability of success for user 2 $p_{s,2}\geq90$. For the considered scenario, $T_\text{int}^*=13$ time slots, implying that $p_{s,2}\geq90$ cannot be guaranteed for $\alpha\gtrsim 1/T_\text{int}^*=0.076$. Conversely, the optimal values of $K$ and $N$ with \gls{noma} increase with $\alpha$. Furthermore, $S_1\geq 0.75$ cannot be guaranteed with \gls{noma} for $\alpha>0.032$. 

\section{Conclusions}\label{sec:concl}

In this paper, we investigated orthogonal and non-orthogonal slicing mechanisms for heterogeneous services in the \gls{ran}. Our analyses and results highlighted the different trade-offs and achievable performance of \gls{oma} and \gls{noma}. Despite the fact that we assumed a conservative collision model for \gls{noma}, which directly impacts the prospects of packet decoding and, hence, the timeliness of the intermittent users, the presented Pareto frontiers for \gls{oma} and \gls{noma} are closely similar.
Thus, \gls{noma} is expected to outperform \gls{oma} in a scenario with capture.
On the other hand, our evaluations showed that finding the optimal configuration with \gls{noma} depends on the activity pattern of the intermittent user, to which \gls{oma} is insensitive.
Our future work includes the analysis and design of slicing mechanisms in scenarios with multiple broadband and intermittent users, as well as capture.
\bibliographystyle{IEEEtran}
\bibliography{IEEEabrv,bib}
\end{document}